# Study of Gas-Fluidization Dynamics with Laser-Polarized $^{129}$Xe

R. Wang[*,+], M. S. Rosen[*], D. Candela[#], R. W. Mair[*,+], and R. L. Walsworth[*]

[*] Harvard-Smithsonian Center for Astrophysics, Cambridge, MA 02138, USA.

[+] Massachusetts Institute of Technology, Dept. of Nuclear Engineering, Cambridge, MA, 02139, USA.

University of Massachusetts, Dept. of Physics, Amherst, MA 01003, USA.

**Corresponding Author:**

Ross Mair

Harvard Smithsonian Center for Astrophysics,

60 Garden St, MS 59,

Cambridge, MA, 02138,

USA

Phone: 1-617-495 7218

Fax: 1-617-496 7690

Email: rmair@cfa.harvard.edu

**ABSTRACT**

We report initial NMR studies of gas dynamics in a particle bed fluidized by laser-polarized xenon ($^{129}$Xe) gas. We have made preliminary measurements of two important characteristics: gas exchange between the bubble and emulsion phases; and the gas velocity distribution in the bed. We used $T_2^*$ contrast to differentiate the bubble and emulsion phases by choosing solid particles with large magnetic susceptibility, in order. Experimental tests demonstrated that this method was successful in eliminating $^{129}$Xe magnetization in the emulsion phase, which enabled us to observe the time-dependence of the bubble magnetization. By employing the pulsed field gradient method, we also measured the gas velocity distribution within the bed. These results clearly show the onset of bubbling and can be used to deduce information about gas and particle motion in the fluidized bed.

## I. INTRODUCTION

Gas fluidization is a process in which solid particles experience fluid-like suspension in an upward flowing gas stream [1,2]. Four different fluidization regimes have been observed, listed in order of increasing gas flow rate: homogeneous fluidization, bubbling fluidization, slugging and pneumatic transport [3]. Homogeneous fluidization indicates the onset of particle suspension, triggered when the weight of the particles is balanced by drag forces from the fluid, including viscous drag, inertial drag and buoyancy. Bubbles, or void spaces with volume much larger than that of a single particle, emerge when the gas flow rate is further increased. Two phases exist in a bubbling bed: one is the bubble phase with almost no particles inside, and the other is the remaining solid-gas mixture with a large particle density, and is known as the emulsion phase. Bubbles rise quickly through the bed, usually at velocities much faster than the upward flow of gas in the emulsion phase, promoting an enhanced circulation and mixing of particles throughout the bed, and quickly relaxing concentration and temperature gradients. Slugging refers to the state where the size of the bubbles approaches that of the container, especially for fast flow through a deep particle bed. Pneumatic transport happens when the flow rate is so high that the gas pushes the particles along with the gas and the particles leave the bed continuously.

Despite the wide application of gas fluidization in industry [1,2], the understanding of the dynamics is far from complete since such a system is difficult to model mathematically, primarily due to the large number of degrees of freedom and inelastic collisions among the particles [4]. A typical fluidized granular system is opaque, resulting in difficulties in probing bed behavior below its surface via light scattering or sound wave techniques [5].

Most commercial fluidized beds operate in the bubbling fluidization regime, in which gas-filled particle-free spaces, the bubbles, emerge at the bottom and expand while rising up along the bed. Bubbles help to agitate the bed to achieve better mixing of particles, but they also provide a shortcut for gas to escape the bed without contacting solid particles [6]. The gas exchange rate in and out of the bubbles measures the efficiency of the contact between the solid and gas phases, which has a significant effect on the operation of the fluidized bed. For example, the reaction rate and yield for a given amount of gas in most chemical reactors is limited by the exchange rate; as is the efficiency of removing moisture in drying processes. A typical scenario for measuring the exchange rate is to inject a pocket of gas of a different species from that in the emulsion phase, and then measure the depletion rate of the gas concentration in the injected bubble [7].

The exchange rate $K$ is defined phenomenologically as follows [3]:

$$-\frac{1}{V_b}\frac{dN_{Ab}}{dt} = K(C_{Ab} - C_{Ae}), \qquad (1)$$

where $V_b$ is the bubble volume; $N_{Ab}$ is the quantity of injected gas species $A$ inside the bubble; $C_{Ab}$ is gas $A$ concentration in the bubble and $C_{Ae}$ is gas $A$ concentration in the emulsion phase. If laser-polarized gas is injected into an emulsion that previously had zero spin polarization, and changes in the bubble volume are ignored, Eqn. (1) can be rewritten as

$$-\frac{dP_b}{dt} = K(P_b - P_e), \qquad (2)$$

where $P_b$ and $P_e$ are the spin polarization in the bubble and emulsion, respectively. The assumption of a constant bubble volume is valid under our experimental conditions [3]. It is therefore possible to measure the exchange rate by monitoring the time-dependence of the spin polarization in the bubble phase.

We used NMR spectroscopy and imaging with laser-polarized $^{129}$Xe as the fluidizing gas to experimentally probe the gas dynamics in a fluidized bed. Previous NMR studies of granular systems have concentrated on the dynamics of the solid particles [8-12]. The $^1$H spins in certain particles have a high signal to noise ratio but are limited when studying gas phase dynamics, since they convey no direct information about the gas flow. To address the common difficulties of low SNR in gas-phase NMR, we employed the spin-exchange optical pumping method [13] to enhance the nuclear spin polarization of $^{129}$Xe gas by ~ 3 orders of magnitude. We report initial results measuring bed behavior at different fluidization regimes regulated by a controllable gas flow rate, which allows us, for the first time, to non-invasively probe bubbles in a fluidized bed, and measure the bubble-emulsion exchange rate.

## II. EXPERIMENTAL SETUP

The experimental apparatus was derived from a setup used previously for the flow of laser-polarized xenon through reservoir rocks [14]. Briefly, xenon gas (26.4% abundance of $^{129}$Xe) was spin-polarized in a glass cell which contained a small amount of Rb metal and a total gas pressure of ~ 4 bar, with ~ 92% xenon and the remainder $N_2$. We heated the cell to $130°$C and induced spin polarization in the resultant Rb vapor via optical pumping at ~ 795 nm, using ~ 60 W of broad-spectrum (~ 2.5 nm) light provided by a fiber-coupled laser diode array [15]. In about 5 minutes of optical pumping, Rb-Xe collisions boost the $^{129}$Xe spin polarization to ~ 1%. The polarized gas then moved through 1/8" I.D. Teflon tubing before flowing through the experimental gas-fluidized bed, and on to a vacuum pump located at the end of the flow path. The gas flow rate was regulated by a mass flow controller, which was capable of providing steady flows ranging from 10 to 1000 cm$^3$/s, placed just before the vacuum pump. We operated in continuous flow mode, where the gas moved continuously from the supply bottles, through the polarization chamber and then the particle bed, and finally through the mass flow controller to the vacuum pump. Xenon gas pressure in the bed was ~ 2.5 bars, due to pressure loss during delivery.

The fluidized bed system used in the experiments consisted of an 8 mm I.D. cylindrical Pyrex column, a windbox and two gas diffusers, which are glass fiber filters with a pore size of 2 µm. The windbox was connected to the fluidization chamber, its large volume providing a buffering space to non-uniform flow patterns as the gas flow direction changes. On top of the windbox was the first gas diffuser, which ensured that the upward flow of gas was homogeneous in cross-section through the particle-holding column, located above. The second diffuser covered the column to stop the particles from escaping out the top. The whole system was assembled with non-magnetic materials so that $^{129}$Xe spin depolarization was minimal during gas delivery. We placed the apparatus in a 4.7 T horizontal bore magnet, interfaced to a Bruker Avance-based NMR console, and we employed a home-built solenoid RF coil for $^{129}$Xe observation at 55.4 MHz.

### III. MEASUREMENT OF GAS EXCHANGE BETWEEN BUBBLE AND EMULSION PHASES

Gas exchange between the bubble and emulsion phases happens in two distinct ways. The coherent penetrating upward flow of gas through the bubble provides the first mechanism for inter-phase exchange, and predominates with smaller bubbles and denser or larger particles [3]. The second source of exchange is the random diffusion of gas molecules through the boundary of the bubble, which is more significant in the case of large bubbles or when a highly diffusive gas species is used.

In order to measure the exchange rate, we required a contrast modality so that the two phases could be clearly differentiated. The obvious difference between the bubble and emulsion is the concentration of solid particles: the emulsion has a large particle density - the total volume of particles is ~ 60% - while more than 99% of the bubble volume is occupied by gas [3]. When placed in a magnetic field of 4.7 Tesla, xenon gas spins in the emulsion phase experience a much larger field inhomogeneity than those in bubbles due to the large susceptibility contrast between the gas and solid phases. Moreover, gas bubbles are almost spherical in shape, and so the resulting field inside the bubble will have a higher homogeneity than that in the emulsion phase. The NMR spectral line from the bubble should therefore be narrower than that from the emulsion, providing a contrast mechanism by which measurement of the exchange rate between the phases is possible.

Fig. 1 shows xenon spectra measured while the polarized gas flows through a bed of alumina particles of average size 50 µm, at four different gas flow rates: 30, 50, 100 and 190 sccm, and at a gas pressure of ~ 2.5 bar. The narrow peak with largest amplitude is due to free gas beyond the bed, which was away from the magnet isocenter, and was therefore frequency-shifted. The broad peak (~ 1.2 kHz FWHM) overlapping the free gas peak is from the emulsion, its width the result of the large field gradients in interstitial spaces. A second broad peak with roughly the same width but shifted 2.6 kHz away was identified to be due to adsorption of xenon onto the particles. (We also performed spectral measurements on a glass cell containing ~ 4 bar of xenon filling the interstitial spaces of a static alumina particle pack. Only two broad peaks were present in this spectrum, with a separation of 1.3 kHz between them, in agreement with previous observations that the adsorption shift is highly related to the interstitial gas pressure [17, 18]. )

We identified the narrow peak of small amplitude, located on top of the emulsion peak, as the bubble phase. This peak increased in amplitude as the gas flow rate was increased from 30 to 190 sccm, which is consistent with known behavior that more bubbles arise and their diameter becomes bigger as the gas flow rate increases [16]. The $T_2^*$ contrast, clearly demonstrated by the emulsion peak being over an order of magnitude broader than the bubble peak, allowed us to differentiate the two phases unambiguously using NMR methods.

We used a stimulated echo sequence to eliminate the emulsion phase $^{129}$Xe polarization, leaving only the bubble peak in which we could observe the time-dependence of the bubble magnetization. The first 90° hard RF pulse flipped spins in both phases non-selectively, before a delay time $\tau_1$ (for the alumina bed, $\tau_1$ was chosen to be 1 ms, which is 3 times $T_2^*$ of the emulsion gas but less than that of the bubble gas) after which only magnetization in the bubble is left. The second 90° RF pulse rotated the bubble magnetization back to the longitudinal direction for storage, benefiting from the long $T_1$ spin polarization lifetime. Gas exchange between the bubble and emulsion phases happened during the subsequent delay $\tau_M$. The last 90° RF pulse then turned the resultant magnetization back to the transverse plane for FID detection. Phase cycling was applied to eliminate the stimulated echo after the third 90° RF pulse. The measured spectrum, for $\tau_M$ = 1 ms to avoid gas exchange, is shown expanded near the bubble peak, in Fig. 2. Both emulsion and adsorption peaks disappeared, demonstrating that the sequence worked effectively in suppressing the magnetization in the emulsion phase.

To measure the exchange rate between the emulsion and bubble phases, we observed the variation in the amplitude of the bubble peak, after the emulsion magnetization had been suppressed, as a function of $\tau_M$. The stimulated echo sequence was used with a series of increasing values of $\tau_M$. The result of this measurement is shown in Fig. 3. From this preliminary data, a critical time $\tau_M$ ~ 0.5 s is evident for polarized $^{129}$Xe gas exchange. The gas exchange time for bubbles sized around 1 mm (estimated for our experiments) has been predicted to be ~ 0.1 s [16]. Potential systematic problems with our preliminary gas exchange measurement include polarized gas from below the bed entering during the exchange time $\tau_M$, and bubbles leaving during $\tau_M$.

## IV. GAS VELOCITY MEASUREMENT

We also measured the velocity of gas flowing in the fluidized bed with the Pulsed Field Gradient Stimulated Echo method [19,20]. Seven different gas flow rates were used to observe the effect of flow rate changes on gas velocity distribution. Glass beads of 50 μm diameter were used in this measurement for better signal strength since the lower magnetic susceptibility of glass gave $^{129}$Xe spectral peaks that were an order of magnitude narrower than with alumina particles. The results of the measurement are shown in Fig. 4. When the gas flow rate was below 30 sccm, the bed was in the homogeneous fluidization regime, in which previous measurements with other methods show that the movement of particles is minimal and gas percolates through the interstitial spaces in the laminar flow regime. This is verified by the results shown in Fig. 4a, where the average velocity increases with gas flow rate but the broadness of the distribution, a measure of random dispersion, is independent of flow rate.

The movement of solid particles greatly affects the gas flow paths, and therefore increases gas dispersivity, as shown in Fig. 4b, in which all the gas flow rates were above 30 sccm and the bed was in the bubbling fluidization regime. The velocity distribution corresponding to 30 sccm is also included for comparison. The width of the peaks increased with the gas flow rate, indicating more random gas flow patterns related to bubble-agitated particle motion. Surprisingly, the average gas velocity decreased at higher gas flow rates in this regime. We believe the reason is that the bubble velocity was larger than the maximum velocity detectable with this method. A larger portion of gas entered the bed in the form of bubbles, during the flow encode time $\Delta$, at higher gas flow rates, and left the bed without being detected, resulting in the observed decreased average gas velocity.

## V. CONCLUSIONS

We performed preliminary measurements of polarized $^{129}$Xe gas exchange between the bubble and emulsion phases, and the gas velocity distribution in a gas-fluidized bed. We applied non-invasive NMR methods so that the fluidization operation was not perturbed by intrusive probe particles, as have been used in earlier measurements [16]. To provide NMR contrast between the bubble and emulsion phases, we exploited the order of magnitude difference in $^{129}$Xe $T_2$* in these two phases. The velocity distribution measurements clearly show the transition from homogeneous to bubbling fluidization.

## ACKNOWLEDGEMENTS

We thank David Cory for technical assistance and useful discussions. We acknowledge support by NSF grant CTS-0310006, NASA grant NAG9-1489 and the Smithsonian Institution.

# REFERENCES

1. R. M. Nedderman, Statics and Kinematics of Granular Materials, Cambridge University Press, Cambridge (1992).
2. B. J. Ennis, J. Green and R. Davies, The legacy of neglect in the U.S, *Chem. Eng. Prog.*, **90,** 32-43 (1994).
3. D. Kunii, O. Levenspiel, Fluidization Engineering, Butterworth-Heinemann, Boston (1991).
4. R. Jackson, The Dynamics of Fluidized Particles, Cambridge University Press, Cambridge (2000).
5. J.-Z. Xue, E. Herbolzheimer, M. A. Rutgers, W. B. Russel, and P. M. Chaikin, Diffusion, Dispersion, and Settling of Hard Spheres, *Phys. Rev. Lett*, **69**, 11-14 (1992).
6. D. Geldart, Gas Fluidization Technology, John Wiley & Sons Ltd, New York (1986).
7. D. J. Patil, M. S. Annaland, and J.A.M. Kuipers, Gas Dispersion and Bubble-to-Emulsion Phase Mass Exchange in a Gas-Solid Bubbling Fluidized Bed: A Computational and Experimental Study, *International Journal of Chemical Reactor Engineering*, **1**, A44 (2003).
8. M. Nakagawa, S. A. Altobelli, A. Caprihan, E. Fukushima and E. K. Jeong, Noninvasive Measurements of Granular Flows by Magnetic Resonance Imaging. *Exp. Fluids*, **16**, 54-60 (1993).
9. E. E. Ehrichs, H. M. Jaeger, G. S. Karczmar, J. B. Knight, Y. V. Kuperman and S. R. Nagel, Granular Convection Observed by Magnetic Resonance Imaging. *Science*, **267**, 1632-1634 (1995).
10. J. D. Seymour, A. Caprihan, S. A. Altobelli and E. Fukushima, Pulsed Gradient Spin Echo Nuclear Magnetic Resonance Imaging of Diffusion in Granular Flow. *Phys. Rev. Lett.* **84**, 266-269 (2000).
11. X. Yang, C. Huan, D. Candela, R. W. Mair and R. L. Walsworth, Measurements of Grain Motion in a Dense, Three Dimensional Granular Fluid. *Phys. Rev. Lett*., **88**, 044301 (2002).
12. R. Savelsberg, D. E. Demco, B. Blumich, and S. Stapf, Particle motion in gas-fluidized granular systems by pulsed-field gradient nuclear magnetic resonance, *Phys. Rev. E*, **65**, 020301 (2002).
13. T. G. Walker and W. Happer, Spin-exchange optical pumping of noble-gas nuclei, *Rev. Mod. Phys*., **69**, 629-642 (1997).
14. R. Wang, R. W. Mair, M. S. Rosen, D. G. Cory and R. L. Walsworth, Simultaneous Measurement of Rock Permeability and Effective Porosity using Laser-Polarized Noble Gas NMR, *Phys. Rev. E*, in press (2004).
15. M. S. Rosen, T. E. Chupp, K. P. Coulter, R. C. Welsh and S. D. Swanson, Polarized Xe-129 optical pumping/spin exchange and delivery system for magnetic resonance spectroscopy and imaging studies, *Rev. Sci. Instrum.,* **70**, 1549-1552 (1999).
16. J. F. Davidson, D. Harrison, Fluidization, Academic Press, London (1972).
17. C. Pak, S. J. Cho, J, Y. Lee, and R. Ryoo, Preparation of Iridium Clusters in Zeolite Y via Cation Exchange: EXAFS, Xenon Adsorption, and $^{129}$Xe NMR Studies, *J. Catalysis*, **149**, 61-69 (1994).

18. A. Bifone, T. Pietrass, J. Kritzenberger, and A. Pines, Surface Study of Supported Metal Particles by $^{129}$Xe NMR, *Phys. Rev. Lett.*, **74**, 16-19 (1995).
19. J. E. Tanner and E. O. Stejskal, Restricted Self-Diffusion of Protons in Colloidal Systems by Pulsed-Gradient Spin-Echo Method, *J. Chem. Phys.* **49**, 1768-1777 (1968).
20. P. T. Callaghan, Principles of Nuclear Magnetic Resonance Microscopy. Oxford University Press, Oxford (1991).

# FIGURE CAPTIONS

**Figure 1**. $^{129}$Xe spectra in an alumina bead pack, measured at four different gas flow rates: 30, 50, 100 and 190 sccm. The narrow peak in the circle is from the bubble phase.

**Figure 2**. $^{129}$Xe spectrum from the stimulated echo sequence with $T_2^*$ contrast to eliminate the emulsion phase signal. The section containing the bubble peak is shown magnified.

**Figure 3**. Preliminary measurement of the gas bubble-emulsion phase exchange time, using the stimulated echo sequence for $T_2^*$ constrast. The integration of the bubble peak is shown as a function of the exchange time $\tau_M$.

**Figure 4**. Xenon gas velocity distributions measured in two fluidization regimes, for 50 µm glass beads. Velocity spectra were measured by the pulsed field gradient stimulated echo technique, in which the gradient pulse duration, $\delta = 1$ ms, the flow encode time $\Delta = 10 \sim 1000$ ms and the maximum gradient pulse strength was 20 G/cm. a). Four different gas flow rates: 10, 16, 21 and 30 sccm were used, all of which ensured the particle bed was in the homogeneous fluidization regime. b). Similar measurements at three higher gas flow rates: 40, 50 and 75, corresponding to the bubbling fluidization regime. Also included is the data for 30 sccm, the transition point between homogeneous and bubbling fluidization.

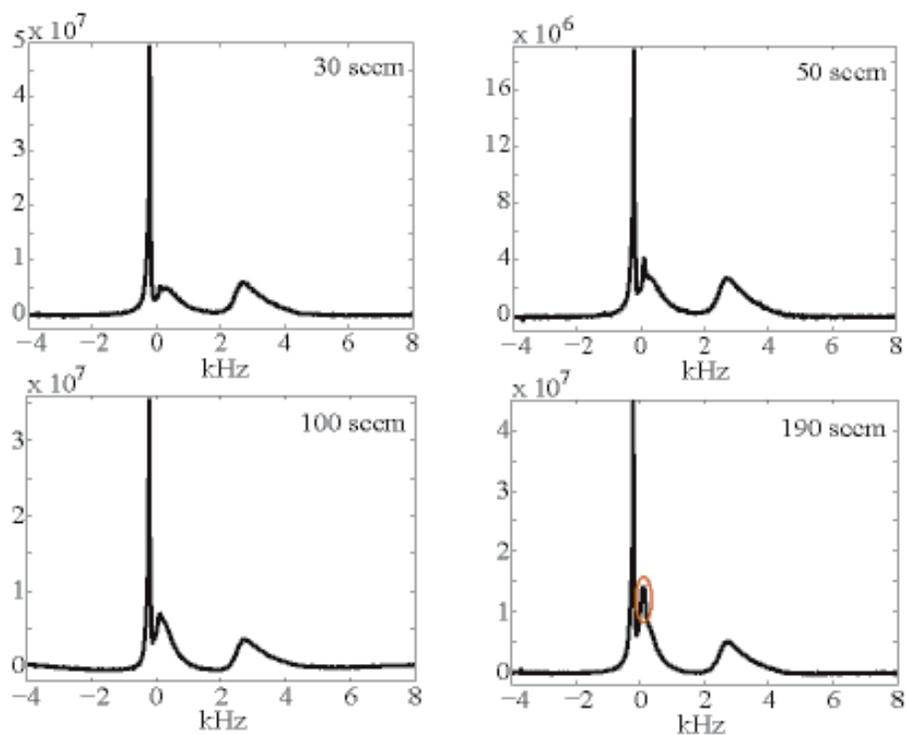

**Figure 1**

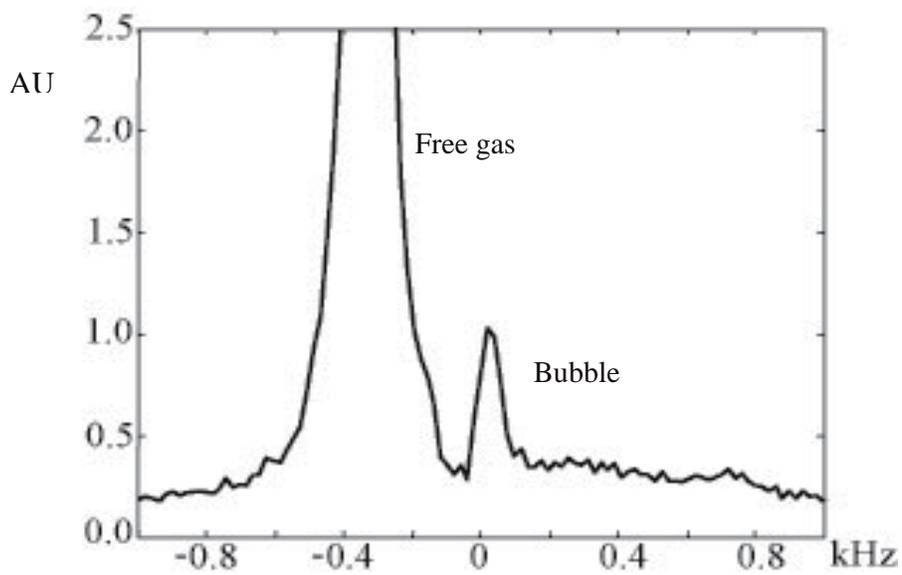

**Figure 2**

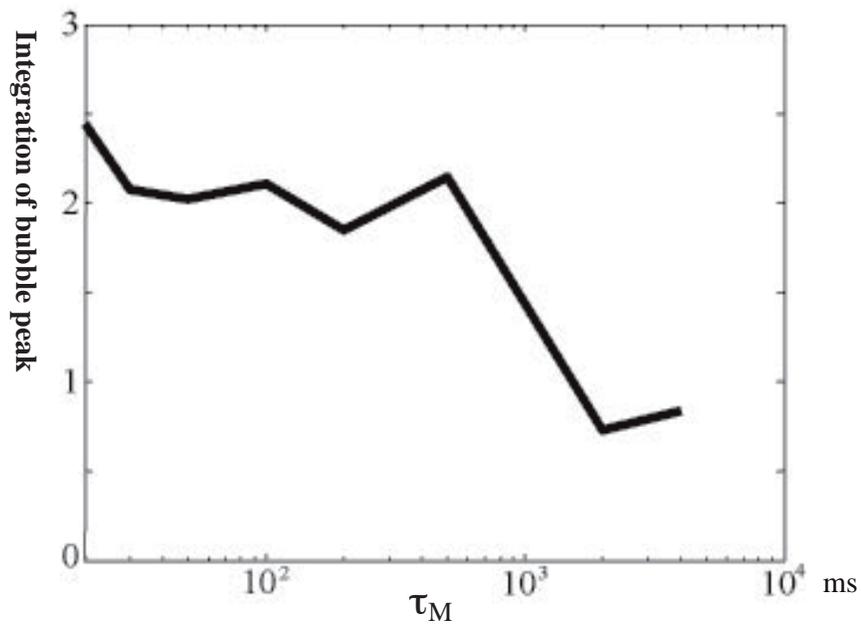

**Figure 3**

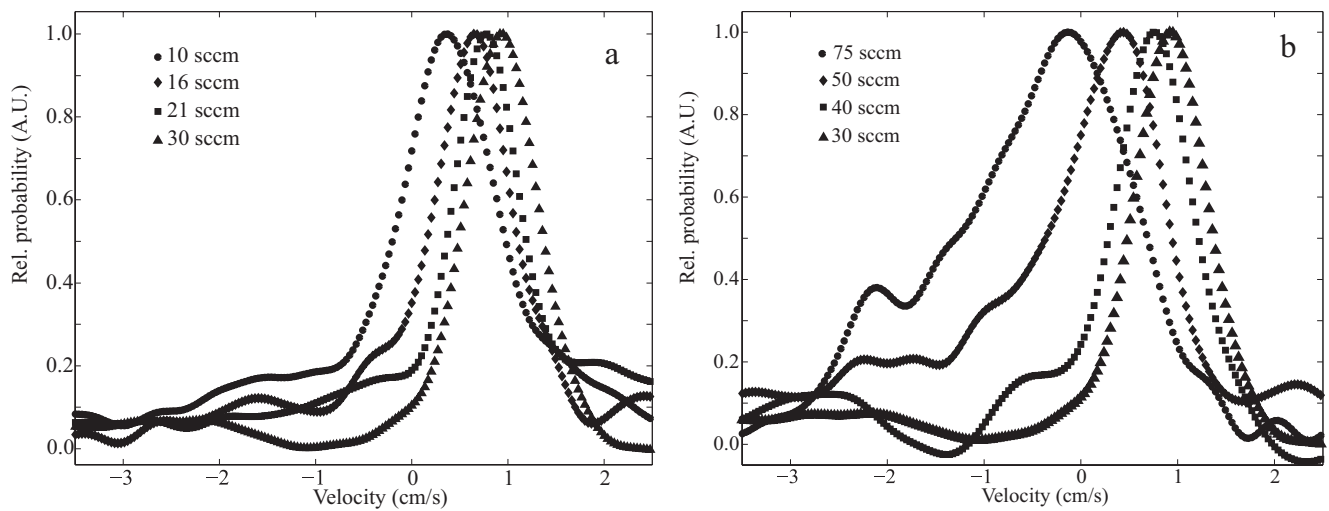

**Figure 4**